\documentstyle[sprocl,twoside,epsf]{article}
\def\Journal#1#2#3#4{{#1} {\bf #2}, #3 (#4)}
\def\PRL{\em Phys. Rev. Lett.}
\def\PRA{{\em Phys. Rev.} A}

\def\PLA{{\em Phys. Lett.} A}

\def\AJP{\em Am. J. Phys.}
\def\RMP{\em Rev. Mod. Phys.}
\def\bx{{\mbox{\boldmath $x$}}}
\def\bk{{\mbox{\boldmath $k$}}}

\def\bq{{\mbox{\boldmath $q$}}}
\def\br{{\mbox{\boldmath $r$}}}
\def\bb{{\mbox{\boldmath $b$}}}
\def\bbn{{\mbox{\boldmath $\scriptstyle n$}}}
\def\bbx{{\mbox{\boldmath $\scriptstyle x$}}}
\def\bbk{{\mbox{\boldmath $\scriptstyle k$}}}
\def\bbb{{\mbox{\boldmath $\scriptstyle b$}}}
\def\bbq{{\mbox{\boldmath $\scriptstyle q$}}}
\def\nab{{\mbox{\boldmath $\nabla$}}}

\def\citebk#1{\hspace{0.9mm}\raisebox{-1.85mm}[0mm][0mm]
  {\Large\cite{#1}}\hspace{-0.1mm}}

\begin{document}
\sloppy

\title{TWO-POTENTIAL APPROACH TO MULTI-DIMENSIONAL TUNNELING}

\author{S.A. GURVITZ}

\address{Department of Particle Physics,
Weizmann Institute of Science,\\ Rehovot~76100, Israel} 

\maketitle
\abstracts{We consider tunneling to the continuum 
in a multi-dimensional potential. 
It is demonstrate that this problem can be treated as two
separate problems: a) a bound state and b) 
a non-resonance scattering problem, by a proper splitting of the
potential into two components. Finally we obtain 
the resonance energy and the partial tunneling widths
in terms of the bound and the scattering state wave functions. 
This result can be used in a variety of tunneling problems.
As an application we consider the 
ionization of atomic states by an external field. We obtain very
simple analytical expressions for the 
tunneling width and for the angular distribution of tunneling electrons.
It is shown that the angular
spread of electrons in the final state is determined by the
semi-classical traversal tunneling time.}

\vspace*{0.5cm}

\tableofcontents

\newpage
\section{Introduction} 
\quad\quad
This paper is devoted to the memory of Michael Marinov, my friend and
collaborator. Within the scope of his broad scientific activity
Michael paid special attention to different aspects of
quantum mechanical tunneling, as for instance
to the coherent and quantum oscillation effects 
in tunneling processes,\cite{mar1,mar2} and to the traversal
tunneling time.\cite{mar,brach} 
In particular Michael was interested in a semi-classical description of
multi-dimensional tunneling. Our frequent discussions of this subject
were very illuminating and helpful to me, especially in relation with
the two-potential approach to multi-dimensional tunneling, which I shall
present below.

It is well known that propagation through classically
forbidden domains has been extensively studied
since the early days of quantum mechanics. Yet, its
exact treatment still remains very complicated and often not
practical, in particular for the multi-dimensional case. 
The treatment of the tunneling problem can be essentially simplified
by reducing it to two separate problems: a bound state plus a
non-resonance scattering state. This can be done consistently in the
two-potential approach (TPA).\cite{gk}$^-$\cite{g1} This approach
provides better physical insight than other
existing approximation methods, and it is  simple and very accurate.
It was originally derived by us for 
the one-dimensional case. Here we present an extension of TPA
to many degrees of freedom and its application to ionization of
atomic states.

\section{General formalism}
\quad\quad
Consider a quantum system in a multi-dimensional space, 
described by the Schr\"odinger equation (we have adapted units where $\hbar=1$) 
\begin{equation}
i{\partial\over\partial t}|\Psi (t)\rangle =[K+V(\bx )]|\Psi (t)\rangle , 
\label{n1}
\end{equation}
where $\bx=\{x_1,x_2,\ldots ,x_n\}$, 
$K=\nab^2/2m$ is the kinetic energy term
and $V(\bx )$ is a multi-dimensional   
potential. The latter contains a barrier that divides 
the entire space into two domains (the ``inner'' and the ``outer'' one),  
such that the classical motion of the system in the inner domain 
is confined. Yet this does not hold for the quantum mechanical motion 
of this system. It is well known that any quantum state with a positive 
energy, localized
initially  inside the inner domain, tunnels to
continuum (the outer domain) through the barrier.  
If the time-dependence of this process is dominated by its
exponential component,
we have a quasi-stationary (resonance) state. 
Below we describe a simple and consistent procedure for treatment of
such multi-dimensional quasi-stationary states, which represents 
a generalization of the TPA, developed in 
Refs.\,\protect\citebk{gk}--\protect\citebk{g1} 
for one-dimensional case.  

Let us first introduce a hyper-surface $\Sigma$ inside the potential 
barrier, which separates the whole space into the inner and 
the outer regions, as shown schematically in Fig.~1.
\begin{figure}[ht] 
\begin{center} 
\leavevmode 
\epsfxsize=8cm 
\epsffile{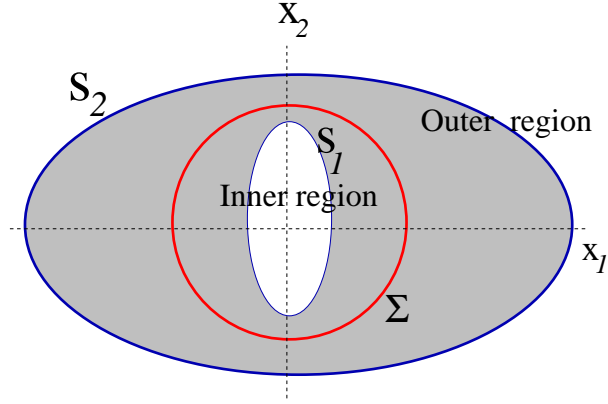} 
\caption{An example of the separation between the inner and the outer  
regions for a two-dimensional potential. $S_1$ and $S_2$ denote the  
equipotential lines for the energy $E_0$, Eq.~(\ref{b1}). The shaded space
correspond   to the classically forbidden region. The separation line $\Sigma$
is   taken as a circle. 
\label{fig:fig1}} 
\end{center} 
\end{figure} 
Respectively, the potential $V(\bx )$ can be represented 
as a sum of two components, $V(\bx )=U(\bx )+W(\bx )$. The first component 
$U(\bx )$ is defined in
such  a way that $U(\bx )=V(\bx )$ in the inner region, and 
$U(\bx )=U_0$ in the outer region, where $U_0$ is the minimal 
value of  $V(\bx )$ on the hyper-surface, $U_0=$min $V(\bx\in\Sigma$).
Respectively,  the potential $W(\bx )=0$ in the inner region, and 
$W(\bx )=V(\bx )-U_0$ in the outer region. 

We start with the bound state $\Phi_0(\bx )$ of the ``inner'' potential,
\begin{equation} 
(K+U)|\Phi _0\rangle =E_0|\Phi_0\rangle\, ,
\label{b1} 
\end{equation}  
where $E_0< U_0$. The potential $W(\bx )$ is switched on at $t=0$. Then the
state $|\Phi_0\rangle $ becomes the wave packet 
\begin{equation}
\Psi(\bx ,t)=b_0(t)\Phi_0(\bx )e^{-iE_0t}+\int
b_{\bbk}(t)\Phi_{\bbk}(\bx )e^{-iE_{\bbk}t}\frac{d\bk}{(2\pi)^n}
\label{b2}
\end{equation}
where $b_0(t), b_{\bbk}(t)$ are the probability amplitudes of finding the
system in the corresponding eigenstates $|\Phi_{0,\bbk}\rangle$ of the
Hamiltonian $H_0=K+U$. For simplicity we assume that the potential $U(\bx )$
contains only one bound state $|\Phi_0\rangle $ corresponding to the energy
$E_0$, Eq.~(\ref{b1}). 
The amplitudes $b_0(t),\, b_{\bbk}(t)$ can be found from the Schr\"odinger 
equation $i\partial_t|\Psi\rangle =(H_0+W)|\Psi\rangle $, supplemented with 
the initial conditions: $b_0(t)=1$, $b_{\bbk}(0)=0$. 

If the state $|\Psi (t)\rangle $ is a metastable one,  
the probability of finding the system inside the inner region drops down
exponentially:  $|b_0(t)|^2\sim e^{-\Gamma t}$, where $\Gamma$ is the width of 
this state. This implies that $\Gamma$ is determined by the exponential 
component of the amplitude $b_0(t)$. The latter corresponds to 
a pole in the complex $E$-plane of the Laplace transformed amplitude 
$\tilde b_0(E)$:   
\begin{equation}
\tilde b_0(E)=\int_0^\infty e^{iEt}\,b_0(t)\,dt
\label{b3}
\end{equation}
The amplitude $\tilde b_0(E)$ can be found from
the Schr\"odinger equation by using the Green's function technique. 
We obtain:\,\cite{g}
\begin{equation}
\tilde b_0(E)=\frac{i}{E-E_0-\langle\Phi_0|W|\Phi_0\rangle-
\langle\Phi_0|W\tilde G(E)W|\Phi_0\rangle}
\label{b4}
\end{equation}
where the Green's function $\tilde G$ is given by
\begin{equation}
\tilde G(E)=G_0(E)(1+\tilde W\tilde G(E)) ,
\label{b5}
\end{equation}
and
\begin{equation} 
G_0(E)=\frac{1-\Lambda}{E+U_0-K-U},\qquad 
\Lambda =|\Phi_0\rangle \langle\Phi_0|\; .
\label{b6}
\end{equation}
Here $\tilde W(\bx )=W(\bx )+U_0$ is the outer component of the potential
$V(\bx )$.  Note that Eqs.~(\ref{b4})- (\ref{b6})
represent the standard perturbation theory, except for 
the ``renormalized'' distortion potential, i.e.
$W(\bx )$ is replaced by $\tilde W(\bx )$ in the Green's function $\tilde G$, 
Eq.~(\ref{b5}).\,\cite{g}

It follows from Eq.~(\ref{b4}) that a pole of $\tilde b_0(E)$ (at 
$E=E_r-i\Gamma /2$), that determines the 
energy $E_r$ and the width  $\Gamma$ of a quasi-stationary state,  
is given by the equation  
\begin{equation} 
E=E_0+\langle\Phi_0|W|\Phi_0\rangle+\langle\Phi_0|W\tilde G(E)W|\Phi_0\rangle\, . 
\label{b7}
\end{equation}
One can easily show\,\cite{g} that Eq.~(\ref{b7}) determines in fact, the
poles in the  total Green's function $G(E)=(E-K-V)^{-1}$ at complex $E$.
Since their position depends only on the potential $V(\bx )$, 
the resonance energy $E_r$ and the width $\Gamma$, given by Eq.~(\ref{b7}) 
are independent of the choice of the separation hyper-surface,
$\Sigma$.  

One can solve Eq.~(\ref{b7}) 
perturbatively by using the standard Born series for the Green's function 
\begin{equation} 
\tilde G=G_0+G_0\tilde W G_0+G_0\tilde W G_0\tilde W G_0+\cdots \ .
\label{b8}
\end{equation}
Substituting Eq.~(\ref{b8}) into Eq.~(\ref{b7}) we obtain the 
corresponding series for $E_r-i\Gamma /2$. Yet, such a series   
converges very slowly. 

For this reason we proposed in Ref.\,\protect\citebk{g} a different expansion 
for $\tilde G$, which converges much faster than the  
Born series. This is achieved by expanding $\tilde G$ 
in powers of the Greens function $G_{\tilde W}$,
\begin{equation}
G_{\tilde W}(E)=\frac{1}{E-K-\tilde W} \,,
\label{b10}
\end{equation}
corresponding to the outer potential $\tilde W$, instead  
of $G_0$, Eqs.~(\ref{b6}), (\ref{b8}), which 
corresponds to the inner potential $U$. 
One then obtains from Eqs.~(\ref{b5}), (\ref{b10})  
\begin{equation}  
\tilde G=G_{\tilde W}+G_{\tilde W}(U-U_0)\tilde G
-\tilde G_{\tilde W}\Lambda 
(1+\tilde W\tilde G )
\label{b9}
\end{equation}

Iterating Eq.~(\ref{b9}) in powers of $G_{\tilde W}$
and then substituting the result into Eq.~(\ref{b7})
we obtain the desirable perturbative expansion for the resonant energy
and the width. Using simple physical reasonings and direct 
evaluation of higher order terms for one dimensional case, 
we demonstrated in Refs.\,\protect\citebk{g}, \protect\citebk{g1} that this expansion converges 
very fast, so that $\tilde G$ is well approximated by the 
first term $G_{\tilde W}$. In this case Eq.~(\ref{b7}) becomes:
\begin{equation} 
E=E_0+\langle\Phi_0|W|\Phi_0\rangle 
+\langle\Phi_0|WG_{\tilde W}(E)W|\Phi_0\rangle,
\label{b11}
\end{equation}  
Solving Eq.~(\ref{b11}) for complex $E=E_r-i\Gamma/2$
we find the resonance energy $E_r$ and 
the width $\Gamma$
with a high accuracy (of order $\Gamma/U_0$).    

Let us assume that that $|\Delta |,\, \Gamma \ll E_0$, where  
$\Delta =E_r-E_0$ is the energy shift. In this case we can solve 
Eq.~(\ref{b11}) iteratively by replacing the argument 
$E$ in the Green's function by $E_0$:
\begin{equation} 
E=E_0+\langle\Phi_0|W|\Phi_0\rangle
+\langle\Phi_0|WG_{\tilde W}(E_0)W|\Phi_0\rangle\,. 
\label{bb11}
\end{equation}
Equation (\ref{bb11}) can be significantly simplified.
Indeed, the Green's function $G_{\tilde W}$ satisfies  
the Schr\"odinger equation 
\begin{equation}  
\left [E_0-K-\tilde W(\bx )\right ] 
G_{\tilde W}(E;\bx ,\bx' )=\delta (\bx -\bx' )\, ,
\label{b12} 
\end{equation}
which can be rewritten as 
\begin{equation} 
W(\bx )G_{\tilde W}(E_0;\bx ,\bx')=
\left ( E_0-U_0-{\nab^2\over 2m}\right )G_{\tilde W}(E_0;\bx ,\bx')
-\delta (\bx -\bx')\ . 
\label{bb12}
\end{equation}
Substituting Eq.~(\ref{bb12}) into Eq.~(\ref{bb11}) we find that 
the corresponding contribution from the kinetic energy term ($\nab^2/2m$)
can be transformed as  
\begin{equation} 
\Phi_0(\bx )\nab^2G_{\tilde W}(E_0;\bx ,\bx')
=\nab (\Phi_0\nab G_{\tilde W})
-\nab (G_{\tilde W}\nab \Phi_0)+G_{\tilde W}\nab^2\Phi_0
\label{b13}
\end{equation}
In addition, the potential $W(\bx )$ is zero in the inner region,
so that the integration over $\bx , \bx'$ in the matrix elements
(\ref{bb11}) takes place only in the outer region, ${\cal R}_{out}$. 
Then using Eq.~(\ref{b13}) and the  
Gauss theorem we find 
\begin{eqnarray} 
\int_{{\cal R}_{out}}\Phi_0(x)\nab^2
G_{\tilde W}(E_0;x,x')d\bx&=&
\int_\Sigma \left [\, G_{\tilde W}\nab_{\bbn}\Phi_0
-\Phi_0\nab_{\bbn} G_{\tilde W}\right ] d\sigma\nonumber\\*[5pt]
&+&\int_{{\cal R}_{out}}G_{\tilde W}\nab^2\Phi_0 d\bx
\label{b14}
\end{eqnarray}
Since the wave function of the (inner) bound state in the outer
region satisfies the Schr\"odinger equation with the potential $U_0$,
the last term in Eq.~(\ref{b14}) can be rewritten as 
\begin{equation}
\int_{{\cal R}_{out}}G_{\tilde W}\nab^2\Phi_0d\bx=
2m(U_0-E_0)\int_{{\cal R}_{out}}G_{\tilde W}\Phi_0d\bx \ , 
\label{bb14}
\end{equation}
Using Eqs.~(\ref{bb12}),  (\ref{b14}), (\ref{bb14})
we then obtain 
\begin{equation} 
\int\Phi_0(x)W(\bx )G_{\tilde W}(E_0;x,x')d\bx=
{1\over 2m}\int\limits_\Sigma \left [\, G_{\tilde W}\nab_{\bbn}\Phi_0
-\Phi_0\nab_{\bbn} G_{\tilde W}\right ] d\sigma
-\Phi_0(\bx' )
\label{bbb14}
\end{equation}
Substituting Eq.~(\ref{bbb14}) into Eq.~(\ref{bb11})
we find that the latter can be rewritten as 
\begin{equation}
E=E_0-{1\over 2m} \int_{\bbx\in\Sigma}\Phi_0(\bx
)\stackrel{\leftrightarrow}\nab_{\bbn}
G_{\tilde W}(E_0;\bx ,\bx')W(\bx')\Phi_0(\bx')d\sigma d\sigma' \ , 
\label{b15}
\end{equation}
where $\stackrel{\leftrightarrow}\nab$ means the gradient on the 
right minus the gradient on the left.

The $\bx'$ integration can be carried out in the same way
by using Eqs.~(\ref{bb12})-(\ref{bb14}). 
Again, the 
$\nab^2\Phi_0(\bx' )$ term cancels as well as the
$\delta$-function contribution. Finally we obtain that
the resonance energy and the width are expressed through a double integral
over  the hyper-surface $\Sigma$
\begin{equation}
E=E_0-{1\over (2m)^2}\int_{\bbx ,\bbx'\in\Sigma}\Phi_0(\bx
)\stackrel{\leftrightarrow}\nab_{\bbn} G_{\tilde W}(E_0;\bx
,\bx')\stackrel{\leftrightarrow}\nab_{\bbn'}\Phi_0(\bx')d\sigma d\sigma'\, .
\label{b16}
\end{equation}

The resonance width $\Gamma$ can be easily
obtained from Eq.~(\ref{b16}) by taking the spectral representation for the
Green's function,  
\begin{equation}
G_{\tilde W}=\int {|\varphi_{\bbk}\rangle\langle \varphi_{\bbk}|\over E_0-E_{\bbk}-i\eta}\,
{d\bk\over (2\pi)^n} \ ,
\label{b17}
\end{equation}   
where $E_{\bbk}=\bk^2/2m$ and $\varphi_{\bbk}$ are the eigenstates
of the ``outer'' Hamiltonian  
corresponding to non-resonance scattering states. 
\begin{equation}
[K+\tilde W(\bx )]\varphi_{\bbk}(\bx )=E_{\bbk}\varphi_{\bbk}(\bx )\, ,
\label{b18}
\end{equation}   
 Using Eqs.~(\ref{b16}), (\ref{b17}) one obtains the total width as an integral
over the partial width $\Gamma_{\bbk}$
\begin{equation}
\Gamma =\left. \int\Gamma_{\bbk }{dk_1\cdots dk_{n-1}\over (2\pi)^{n-1}}
\right |_{|\bbk |=k_0} 
\label{b19}
\end{equation}
where $k_0=\sqrt{2mE_0}$  and 
\begin{equation} 
\Gamma_{\bbk }={1\over 4mk_n}\left |\int_{\bbx\in\Sigma}\Phi_0(\bx)
\stackrel{\leftrightarrow}\nab_{\bbn}\varphi_{\bbk}(\bx)d\sigma
\right |^2_{|\bbk |=k_0} \ .  
\label{b20}
\end{equation}

Equations.~(\ref{b19}) and (\ref{b20})  represent our final result for
the total and partial tunneling widths of the resonance state with 
energy $E_r=E_0$, Eq.~(\ref{b1})\,.\,\footnote{~Equation (\ref{b20}) 
resembles the Bardeen formula\,\cite{ft} for
energy splitting of two degenerate levels in a double-well
potential.}
The latter is the energy of a pure bound state $\Phi_0$ in the inner
potential $U(\bx )$. Both $\Gamma_\bbk$ and $E_0$ are independent of
the hyper-surface $\Sigma$ (up to the terms of the order of
$\Gamma /U_0$), providing that $\Sigma$ is taken far away from
the equipotential surfaces $V(\bx )=E_0$\,.\,\cite{g1}
 
Equations.~(\ref{b1}), (\ref{b19}), (\ref{b20}) considerably simplify the
multi-dimensional tunneling problem, by reducing it to the bound state
problem. The latter can be treated with standard techniques. 
Some examples of applications of our method are given below.  

\section{Ionization of bound state in semi-classical approximation }

\quad\quad 
Consider a three-dimensional potential $u(\br )$, 
which contains a bound state $\phi_0 (\br )$ with the energy
$\epsilon_0<0$. Let us superimpose an additional potential $v(x)$,
depending on the coordinate $x$. We assume that  $v(x)$
ionizes the bound state $\phi_0$, i.e. 
$u(\br )+v(x)<\epsilon_0$ for $x\to\infty$.
We also assume that $u(\br )$ and $v(x)$ do not
overlap, such that  $u(\br )\approx 0$ for $x>x_1$ and
$v(x)\approx 0$ for $x<x_2$, where $x_1<x_2$. Now
taking  the separation hyper-surface  $\Sigma$ as a plane, $x=x_0$
where $x_1< x_0< x_2$, we can identify $u(\br )$ and $v(x)$ with the inner
and the outer potentials, $U$ and $\tilde W$, Eqs.~(\ref{b1}), (\ref{b18}).
Then the bound state wave function $\phi_0(\br )$ coincides 
with the inner wave function $\Phi_0$; Eq.~(\ref{b1}).

Providing that $v(x)$ is a function of $x$ only, the
outer wave function, Eq.~(\ref{b18}) can be written as a product
$\varphi_{\bbk}(\br )=\exp (i\bq \bb)\chi_k(x)$, with $\br =(x,\bb )$,
$\bk=(k,\bq )$ and $\bk^2=2m(u_0+\epsilon_0)$, where 
$u_0=-v(x\to\infty )$. As a result, Eq.~(\ref{b20}) for
the partial width $\Gamma_q$ reads   
\begin{equation}
\Gamma_q=\frac{1}{4m\bar k}\left |\int e^{i\bbq\bbb}\left [
\phi_0(x,\bb )\frac{\partial}{\partial x}\chi_{\bar k}(x)-
\chi_{\bar k}(x)\frac{\partial}{\partial x}
\phi_0(x,\bb )\right ] d\,^2 b \right |_{x=x_0}^2\ ,
\label{a3}
\end{equation}
where $\bar k =\sqrt{2m(u_0+\epsilon_0)-\bq^2}$. 
Respectively, the total width, Eq.~(\ref{b19}), is given by    
\begin{equation}
\Gamma =\int\Gamma_q\, \frac{d\,^2q}{(2\pi)^2}
\label{a2}
\end{equation}

Since $\chi_{\bar k}(x)$ is a non-resonant wave function,
it falls down exponentially inside the barrier and
can be evaluated by using the standard semi-classical 
approximation.\cite{shiff}
One can write
\begin{equation}
\chi_{\bar k}(x)=\sqrt{\frac{\bar k}{|\tilde p_q(x)|}}\,
e^{-\!\int_x^{x_f}|\tilde p_q(x')|dx'}\ .
\label{a5}
\end{equation}
Here  $\tilde p_q(x)=\sqrt{2m[\epsilon_0-v(x)]-q^2}$, and  
$x_{f}$ is the outer classical turning point, $\tilde p_q(x_f)=0$. 

The inner wave-function describes the bound state and therefore
it decreases exponentially under the barrier, as well  
\begin{equation}
\phi_0(x,\bb )\sim
e^{-\!\int_{x_i}^{(x^2+\bbb^2)^{1/2}}|\tilde p(x',\bbb )|dx'}\, .
\label{a4}
\end{equation}
Here $|\tilde p(x,\bb )|=\sqrt{2m[\epsilon_0-u(x,\bb )]}$, and $x_i$ is the
inner classical  turning point. It follows from Eq.~(\ref{a4})
that $\phi_0(x,\bb )$ decreases very quickly with $b$ so that  
only small $b$ contribute in the integral (\ref{a3}). Thus, 
we can approximate the inner wave function and its derivative as  
\begin{eqnarray}
&&\phi_0(x,\bb )\simeq \frac{x_i}{x}\,\phi_0(x_i,0)\,
e^{-\!\int_{x_i}^{x}|\tilde p(x',0)|dx'-|\tilde p(x,0)|\bbb^2/2x}\; ,
\label{a6a}\\
\noalign{\vskip10pt}
&&\frac{\partial \phi_0(x,\bb)}{\partial x}
\simeq -|\tilde p(x,0)|\,\phi_0(x,\bb )\; .
\label{a6b}
\end{eqnarray}
Substituting Eqs.~(\ref{a5}), (\ref{a6a}), (\ref{a6b}) into Eq.~(\ref{a3}) we 
can perform the $\bb$-integration by using 
\begin{equation}
\int e^{i\bbq\bbb }\,
\phi_0(x_0,\bb )d\,^2b =\frac{2\pi x_i}{|\tilde p(x_0,0)|}\,
\phi_0(x_0,0)\,e^{-\!\int_{x_i}^{x_0}|
\tilde p(x',0)|dx'-x_0\bbq^2/2|\tilde p(x_0,0)|}\, .
\label{a7}
\end{equation}
Since $x_0\gg x_i$ the exponent in Eq.~(\ref{a7}) can be
approximated as 
\begin{equation}
\int_{x_i}^{x_0}|\tilde p(x',0)|dx'+\frac{x_0q^2}{2|\tilde p(x_0,0)|}\approx  
\int_{x_i}^{x_0}[\tilde p^2(x',0)+\bq^2]^{1/2}dx'
\label{a8}
\end{equation}

Now using Eqs.~(\ref{a7}), (\ref{a8}), and $\partial \chi_{\bar k}/\partial x
\simeq |\tilde p_q(x)|\chi_{\bar k}$, we obtain the following formula for the
partial  width,
\begin{equation}
\Gamma_q=\frac{4\pi^2x_i^2\phi_0^2(x_i,0)}{m(2m|\epsilon_0|)^{1/2}}
e^{-2\int_{x_i}^{x_f}p_q(x)dx}\, ,
\label{a9}
\end{equation}
where 
\begin{equation}
p_q(x)=\sqrt{2m[u(x,0)+v(x)-\epsilon_0]+\bq^2}\, ,
\label{aa9}
\end{equation}
is the absolute value of the imaginary momentum in the 
direction of the external field.  

\section{Partial width and the tunneling time}

\quad\quad
In contrast with the one-dimensional tunneling, the final state in a
multi-dimensional  tunneling is highly degenerate.  Hence, a tunneling
particle can be found in  different final states with the same total energy.
In our case these are all the states with different lateral momenta $\bq$,
leading to the angular distribution of a tunneling particle.
Using our result for the partial width, Eq.~(\ref{a9}), one easily finds that
this distribution is of a Gaussian shape.  Indeed, by expanding the exponent
of Eq.~(\ref{a9}) in powers of $\bq^2$ we obtain   
\begin{equation}
\int_{x_i}^{x_f}p_q(x)dx\simeq 
S(E_0,x_i,x_f)+\frac{q^2}{2m}\int_{x_i}^{x_f}\frac{m
dx}{p(x)} \, ,
\label{a10}
\end{equation}
where $p(x)=p_{q=0}(x)$ and 
\begin{equation}
S(\epsilon_0,x_i,x_f)=\int_{x_i}^{x_f}p(x)dx
\label{aa10}
\end{equation}
is the action in the classically forbidden region. Then Eq.~(\ref{a9})
reads 
\begin{equation}
\Gamma_q=\frac{4\pi^2x_i^2\phi_0^2(x_i,0)}{m(2m|\epsilon_0|)^{1/2}}
e^{-2S(\epsilon_0,x_i,x_f)- \tau q^2/m }\, ,
\label{a12}
\end{equation}
where 
\begin{equation}
\tau =\int_{x_i}^{x_f}\frac{m dx}{|p(x)|}\, ,
\label{a11}
\end{equation}
is the B\"uttiker-Landauer tunneling time.\cite{but} The latter 
is assumed to be an amount of time, which a particle spends 
traversing a tunnel barrier.  It appears, however, that such a quantity 
is not uniquely defined in one-dimensional
tunneling.\cite{hauge,mar} Yet, the situation is different for the
multi-dimensional case. Our result, Eq.~(\ref{a12}), shows 
that  the B\"uttiker-Landauer traversal tunneling time $\tau$ 
is related to a well defined quantity, namely to the lateral spread of a
tunneling particle ($r_0$).\,\cite{brach} Indeed, the latter
corresponds to the inverse average lateral momentum, $r_0=
1/<q>$, where $<q>=\sqrt{m/\tau}$ as obtained from Eq.~(\ref{a12}). 

The Gaussian shape of angular distribution of tunneling particles is a general
phenomenon, which can be understood in the following
way.\cite{brach} It is well known that the most probable tunneling
path corresponds to the minimal action. In our case this path is along 
the direction of the external field. 
Therefore any momentum $\bq$, normal to this direction
raises the action by a term proportional to the corresponding energy 
$\bq^2/2m$. This leads to the Gaussian fall-off of the related probability 
distribution. 

\section{Total width and angular distribution of ionized particles}
\quad\quad
Substituting Eq.~(\ref{a12}) into Eq.~(\ref{a2}) and performing the
$q$-integration, we obtain the following expression for the total width
\begin{equation}
\Gamma
=\frac{\pi\, x_i^2\,\phi_0^2(x_i,0)}{\tau\, (2m|\epsilon_0|)^{1/2}}\,
e^{-2S(E_0,x_i,x_f)}\; .
\label{a13}
\end{equation}
Using this result, Eq.~(\ref{a12}) can be rewritten in a more compact
form as
\begin{equation}
\Gamma_{\bbq}={4\pi\over m}\,\Gamma\,\tau\, e^{-\tau\bbq^2/m}
\label{aa13}
\end{equation}

Let us consider two exemplifying  applications of Eqs.~(\ref{a13}), 
(\ref{aa13}).\\

\noindent
{\em (a) Hydrogen atom}\\

Consider the Hydrogen atom in the ground state under the electric field ${\cal E}$. 
We use the atomic units in which $m=1$,  the ground state energy
$\epsilon_0=-1/2$ and the ground state wave function 
$\phi_0(r)=\exp(-r)/\sqrt{\pi}$.  
In these units the inner and the outer potentials are: $u(r)= -1/r$ and
$v(x)=-{\cal E}x$. Then the action $S$, Eq.~(\ref{aa10}) reads
\begin{equation}
S=\int_{x_i}^{x_f}\sqrt{1-\frac{2}{x}-2{\cal E}x}dx\simeq
\int_{x_i}^{x_f}\sqrt{1-2{\cal E}x}dx-\int_{x_i}^{x_f}\frac{dx}
{x\sqrt{1-2{\cal E}x}} \, .
\label{a14}
\end{equation}
Although the main contribution to the action is coming from large $x$, 
we have keep in (\ref{a14}) also the next term of expansion
(see Ref.\,\protect\citebk{lan}). Performing the integration in Eq.~(\ref{a14}) 
we obtain for the action 
\begin{equation}
S\simeq \frac{1}{3{\cal E}}+\ln\left (\frac{{\cal E}x_i}{2}\right )-
\frac{x_i}{2} \ ,
\label{a15}
\end{equation}
where we assumed that ${\cal E}x_i\ll 1$. Respectively for 
the ``tunneling'' one obtains $\tau =1/{\cal E}$. Using these results we
easily find from Eq.~(\ref{a13}) the well-known result for the ionization width
of the Hydrogen atom\,\cite{lan}
\begin{equation}
\Gamma = {4\over {\cal E}}\,e^{-2/3{\cal E}}\, .
\label{a16}
\end{equation}
Respectively, the angular distribution of ionized electrons, 
Eq.~(\ref{aa13}), is
\begin{equation}
\Gamma_\bbq = {4\pi\over {\cal E}}\, \Gamma e^{-\bbq^2/{\cal E}}\, .
\label{aa16}
\end{equation} 

\noindent
{\em (b) Potential well with short range forces}\\

The bound $s$-state wave function of a short range potential 
at distances larger then the radius of the well $a$ has the
following form 
\begin{equation}
\phi_0(r)={A\sqrt{\kappa}\over r}\,e^{-\kappa r}\, ,
\label{a17}
\end{equation}
where $\kappa =(2|\epsilon_0|)^{1/2}$ and $A$ is a dimensionless constant
depending on the specific form of the well\,\cite{lan}. For instance, if 
$a\ll 1/\kappa$, then $A=(2\pi )^{-1/2}$.  

Similar to the previous case we use the same units but now with $\kappa =1$. 
The action and the tunneling time are evaluated straightforwardly from 
Eqs.~(\ref{aa10}), (\ref{a11}) taking into account that 
the initial turning points are $x_i=a\to 0$. One finds
\begin{eqnarray}
S&=&\int_{0}^{1/2{\cal E}}\sqrt{1-2{\cal E}x}dx=
{1\over 3{\cal E}}
\label{a18}\\*[5pt]
\tau&=&\int_0^{1/2{\cal E}}{dx\over\sqrt{1-2{\cal E}x}}={1\over\ {\cal E}}
\label{a19}
\end{eqnarray}
Using $x_i^2\phi_0^2(x_i,0)=A^2$ one obtains from Eq.~(\ref{a13})
\begin{equation}
\Gamma =\pi A^2\,{\cal E}\,e^{-2/3{\cal E}}\, .
\label{a20}
\end{equation}
This coincides with the result obtained by Demkov and
Drukarev,\cite{lan,druk} using a more complicated technique.

The angular distribution of ionized particles, Eq.~(\ref{aa13}), is 
\begin{equation}
\Gamma_\bbq =4\pi^2 A^2e^{-2/3{\cal E}}e^{-\bbq^2/{\cal E}}\, .
\label{a21}
\end{equation}
The same result was obtained earlier in Ref.~\protect\citebk{brach}.

\section{Summary}

\quad\quad
In this paper we extended the two-potential approach to quantum tunneling 
to a multi-dimensional case. We found simple expressions
for the energy and the partial widths of a quasi-stationary
state, initially localized inside a multi-dimensional potential.
The partial width, which determines the distribution of tunneling
particles in the final state, is given by an integral of  
the bound and non-resonant scattering wave functions
over the hyper-surface taken inside the potential barrier.

We applied our results for the ionization of atomic states by an
external field depending only on one of the coordinates. 
In this case we obtained very simple semi-classical expressions
for the total ionization width and the angular distribution of
tunneling electrons, valid for any initial state. 
We demonstrated that in the case of the Hydrogen atom and
also for a short range potential our general expression
for the total width produces the results,
existing in the literature, but obtained with a more
complicated calculations. 

We also demonstrated that angular distribution of tunneling
electrons in the final state falls off as a Gaussian function
of the lateral momentum. The exponential is determined by
the traversal semi-classical tunneling time.

\section*{Acknowledgments}
\addcontentsline{toc}{section}{\numberline{}Acknowledgments}
I am grateful to M.~Kleber, W.~Nazarewicz, P.~Semmes and S.~Shlomo
for helpful discussions. I would like to acknowledge the kind hospitality of 
Oak Ridge National Laboratory, while parts of this work were being
performed.

\section*{References}
\addcontentsline{toc}{section}{\numberline{}References}

\end{document}